\documentclass[10pt,aps,prb,twocolumn,nofootinbib,floatfix,superscriptaddress]{revtex4-1}

\usepackage{amssymb,amsmath}
\usepackage{subfig}
\usepackage{graphicx}
\usepackage{color}
\usepackage{booktabs, makecell}
\usepackage{siunitx}
\usepackage{listings}
\usepackage[font=small,
   justification=justified,
   format=plain]{caption}
\usepackage{soul}
\graphicspath{{figures/}}

\begin{document}

\newcommand{\1}{{\bf \scriptstyle 1}\!\!{1}}
\newcommand{\I}{{\rm i}}
\newcommand{\p}{\partial}
\newcommand{\D}{^{\dagger}}
\newcommand{\hbe}{\hat{\bf e}}
\newcommand{\bfa}{{\bf a}}
\newcommand{\bx}{{\bf x}}
\newcommand{\hbx}{\hat{\bf x}}
\newcommand{\by}{{\bf y}}
\newcommand{\hby}{\hat{\bf y}}
\newcommand{\br}{{\bf r}}
\newcommand{\hbr}{\hat{\bf r}}
\newcommand{\bj}{{\bf j}}
\newcommand{\bk}{{\bf k}}
\newcommand{\bn}{{\bf n}}
\newcommand{\bv}{{\bf v}}
\newcommand{\bp}{{\bf p}}
\newcommand{\bq}{{\bf q}}
\newcommand{\tp}{\tilde{p}}
\newcommand{\tbp}{\tilde{\bf p}}
\newcommand{\bu}{{\bf u}}
\newcommand{\hbz}{\hat{\bf z}}
\newcommand{\bA}{{\bf A}}
\newcommand{\calA}{\mathcal{A}}
\newcommand{\calB}{\mathcal{B}}
\newcommand{\tC}{\tilde{C}}
\newcommand{\bD}{{\bf D}}
\newcommand{\bE}{{\bf E}}
\newcommand{\calF}{\mathcal{F}}
\newcommand{\bB}{{\bf B}}
\newcommand{\bG}{{\bf G}}
\newcommand{\calG}{\mathcal{G}}
\newcommand{\obG}{\overleftrightarrow{\bf G}}
\newcommand{\bJ}{{\bf J}}
\newcommand{\bK}{{\bf K}}
\newcommand{\bL}{{\bf L}}
\newcommand{\tL}{\tilde{L}}
\newcommand{\bP}{{\bf P}}
\newcommand{\calP}{\mathcal{P}}
\newcommand{\bQ}{{\bf Q}}
\newcommand{\bR}{{\bf R}}
\newcommand{\bS}{{\bf S}}
\newcommand{\bH}{{\bf H}}
\newcommand{\balpha}{\mbox{\boldmath $\alpha$}}
\newcommand{\talpha}{\tilde{\alpha}}
\newcommand{\bsigma}{\mbox{\boldmath $\sigma$}}
\newcommand{\hbeta}{\hat{\mbox{\boldmath $\eta$}}}
\newcommand{\bSigma}{\mbox{\boldmath $\Sigma$}}
\newcommand{\bomega}{\mbox{\boldmath $\omega$}}
\newcommand{\bpi}{\mbox{\boldmath $\pi$}}
\newcommand{\bphi}{\mbox{\boldmath $\phi$}}
\newcommand{\hbphi}{\hat{\mbox{\boldmath $\phi$}}}
\newcommand{\btheta}{\mbox{\boldmath $\theta$}}
\newcommand{\hbtheta}{\hat{\mbox{\boldmath $\theta$}}}
\newcommand{\hbxi}{\hat{\mbox{\boldmath $\xi$}}}
\newcommand{\hbzeta}{\hat{\mbox{\boldmath $\zeta$}}}
\newcommand{\brho}{\mbox{\boldmath $\rho$}}
\newcommand{\bnabla}{\mbox{\boldmath $\nabla$}}
\newcommand{\bmu}{\mbox{\boldmath $\mu$}}
\newcommand{\bepsilon}{\mbox{\boldmath $\epsilon$}}

\newcommand{\iLambda}{{\it \Lambda}}
\newcommand{\cL}{{\cal L}}
\newcommand{\cH}{{\cal H}}
\newcommand{\cU}{{\cal U}}
\newcommand{\cT}{{\cal T}}

\newcommand{\be}{\begin{equation}}
\newcommand{\ee}{\end{equation}}
\newcommand{\bea}{\begin{eqnarray}}
\newcommand{\eea}{\end{eqnarray}}
\newcommand{\beqa}{\begin{eqnarray*}}
\newcommand{\eeqa}{\end{eqnarray*}}
\newcommand{\nn}{\nonumber}
\newcommand{\DD}{\displaystyle}

\newcommand{\ba}{\begin{array}{c}}
\newcommand{\baa}{\begin{array}{cc}}
\newcommand{\baaa}{\begin{array}{ccc}}
\newcommand{\baaaa}{\begin{array}{cccc}}
\newcommand{\ea}{\end{array}}

\newcommand{\bma}{\left[\begin{array}{c}}
\newcommand{\bmaa}{\left[\begin{array}{cc}}
\newcommand{\bmaaa}{\left[\begin{array}{ccc}}
\newcommand{\bmaaaa}{\left[\begin{array}{cccc}}
\newcommand{\ema}{\end{array}\right]}

\definecolor{dkgreen}{rgb}{0,0.6,0}
\definecolor{gray}{rgb}{0.5,0.5,0.5}
\definecolor{mauve}{rgb}{0.58,0,0.82}

\lstset{frame=tb,
  	language=Matlab,
  	aboveskip=3mm,
  	belowskip=3mm,
  	showstringspaces=false,
  	columns=flexible,
  	basicstyle={\small\ttfamily},
  	numbers=none,
  	numberstyle=\tiny\color{gray},
 	keywordstyle=\color{blue},
	commentstyle=\color{dkgreen},
  	stringstyle=\color{mauve},
  	breaklines=true,
  	breakatwhitespace=true
  	tabsize=3
}

\title{Multiferroic Dark Excitonic Mott Insulator in the Breathing-Kagome Lattice Material Nb$_3$Cl$_8$}
\author{Mahtab A. Khan}
\affiliation{NanoScience Technology Center, University of Central Florida, Orlando, FL 32826, USA}
\affiliation{Department of Physics, Federal Urdu University of Arts, Sciences and Technology, Islamabad 44000, Pakistan}

\author{Naseem Ud Din}
\affiliation{Department of Physics, Florida Atlantic University, Boca Raton, FL 33431, USA}

\author{Dmitry Skachkov}
\affiliation{NanoScience Technology Center, University of Central Florida, Orlando, FL 32826, USA}

\author{Dirk R. Englund}
\affiliation{Department of Electrical Engineering and Computer Science,
             Massachusetts Institute of Technology, Cambridge, MA 02139, USA}

\author{Michael N. Leuenberger}
\affiliation{NanoScience Technology Center, University of Central Florida, Orlando, FL 32826, USA}
\affiliation{Department of Physics and College of Optics and Photonics,
             University of Central Florida, Orlando, FL 32826, USA}
\email{leuenberger@ucf.edu}
\date{\today}

\begin{abstract}
Flat electronic bands strongly enhance Coulomb interactions and can stabilize unconventional insulating states. Motivated by the recent discovery of flat bands in breathing Kagome lattices, we use first-principles GW--Bethe--Salpeter theory to investigate the excitonic spectrum of single-layer Nb$_3$Cl$_8$. We find a dark spin-triplet Frenkel exciton whose spectral peak lies at negative energy ($-0.14$~eV) relative to the quasiparticle gap, directly signaling a 
preformed bound state and an excitonic Mott insulating phase potentially stable at room temperature. Bright excitons appear at $0.94$~eV and $1.21$~eV, with ultra-large binding energies of $2.05$~eV and $1.77$~eV. By mapping the low-energy dynamics onto a spin-1 Hubbard model on a triangular lattice, we show that frustrated antiferromagnetic and ferroelectric tendencies naturally emerge. These results identify Nb$_3$Cl$_8$ as a candidate multiferroic dark excitonic insulator, opening a pathway to correlated quantum phases in two dimensions.
\end{abstract}

\maketitle

\section*{Introduction}

Flat bands (FBs), characterized by nearly constant energy across momentum states, lead to highly localized electronic states with large effective masses and suppressed particle hopping. These FBs foster strong electron correlations and enhanced light-matter interactions, giving rise to novel quantum phenomena. Kagome lattices (KLs), composed of corner-sharing triangles forming a hexagonal structure, can host FBs through geometry and interference-driven electron localization. 
\begin{figure*}[!t]\includegraphics[width=\textwidth]{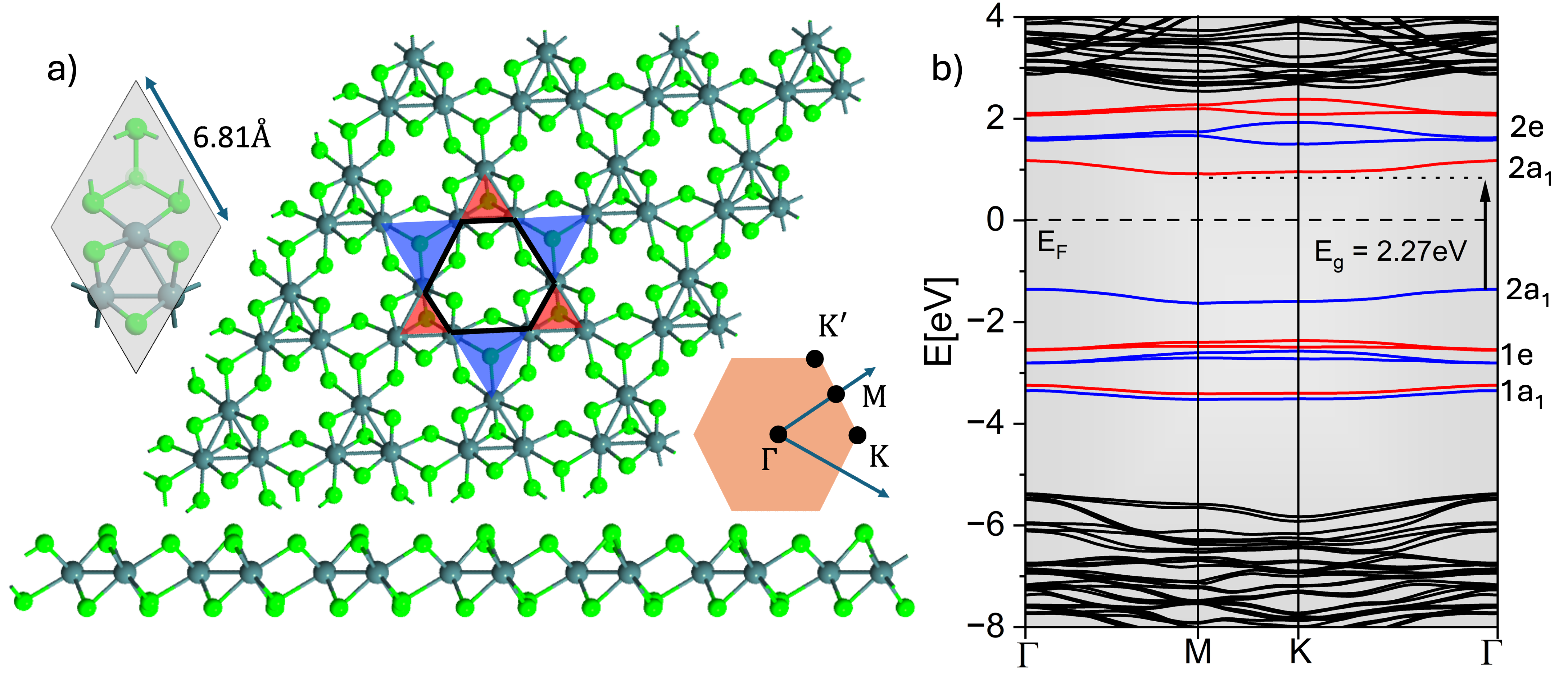}
\centering
\caption{a) Top and side view of Nb$_3$Cl$_8$, shaded grey (orange) region (rhombus) is the unit cell (Brillouin zone) of Nb$_3$Cl$_8$ consisting of 3 Nb (dark green balls) and 8 Cl (light green balls) atoms. Breathing Kagome lattice is formed by Nb atoms creating irregular hexagons (black lines) with two sets of sides of different lengths, resulting in two sets of outer equilateral triangles with different areas (shaded red and blue). SL Nb$_3$Cl$_8$ has C$_{3v}$ symmetry. b)  GW band structure of SL Nb$_3$Cl$_8$, showing substantial increase in the electronic band gap E$_g^{\textrm{GW}}=$2.27 eV as compared with the PBE band gap of E$_g^{\textrm{PBE}}=$0.27 eV, which is shown in the Supplementary Information (SI). Colored lines show flat bands. the blue (red) represents the up (down) spin component, with no preferential direction, as the flat bands are spin-polarized. Black solid lines correspond to the continuum of states with dispersion. Irreducible representations correspondinng to the energy levels at $\Gamma$ point are also shown as a$_1$ and e.}
\label{fig:Nb3Cl8_Structure_Bs}
\end{figure*} 

The breathing Kagome lattice (BKL), characterized by alternating triangles with unequal bond lengths, observed in Nb halides Nb$_3$X$_8$ (such as Nb$_3$Cl$_8$\cite{Sun2022}, Nb$_3$I$_8$,\cite{regmi2022spectroscopic} and Nb$_3$Br$_8$,\cite{PhysRevB.108.L121404} where Nb atoms form a BKL)  introduces a unique geometry that enhances electron localization. In Nb$_3$Cl$_8$, the BKL’s network of corner-sharing triangles (see Fig.~\ref{fig:Nb3Cl8_Structure_Bs}) generates destructive interference in electronic wavefunctions, suppressing kinetic energy and producing dispersionless flat bands (FBs) in momentum space. The breathing distortion breaks inversion symmetry, isolating FBs from dispersive bands and enhancing electron correlations. This suppressed kinetic energy increases Coulomb interactions, enabling exotic quantum phenomena such as strongly bound excitons, magnetism, and potential topological phases. Remarkably, the flat bands in Nb$_3$Cl$_8$ are both distinct see Fig.~\ref{fig:Nb3Cl8_Structure_Bs} (b) and experimentally accessible, in contrast to twisted bilayer semiconductors and traditional Kagome lattices (KLs), where FBs often overlap with a multitude of other bands near the Fermi level or are far away from the Fermi energy, complicating their detection. Moreover, the layered structure of Nb$_3$Cl$_8$ allows it to be thinned down to the monolayer limit,\cite{Sun2022} further enhancing its appeal and versatility in physical sciences, material science, and engineering.

The Exciton Insulator (EI) phase represents a fascinating quantum state of matter where the Coulomb interaction between electrons and holes drives a spontaneous formation of bound excitonic pairs, leading to a macroscopic condensation. \cite{Mott1961, Keldysh1968} The concept of the excitonic insulator was formally introduced by Jérôme, Rice, and Kohn, \cite{Jerome1967} who demonstrated that such a phase can emerge in systems with either a small bandgap (narrow-gap semiconductors) or a band overlap (semimetals), where strong electron-hole interactions induce a collective instability. Rice later extended these ideas to discuss the transition between the excitonic insulator and the electron-hole liquid phases, emphasizing the role of density and screening effects in stabilizing these quantum states.\cite{Rice1970} Advances in experimental and theoretical techniques have revealed the signatures of EI phases in materials like transition metal dichalcogenides (e.g., monolayer TiSe$_2$),\cite{Cercellier2007, Monney2011} topological insulators,\cite{Li2019} and engineered heterostructures.\cite{Fogler2014} Recent studies also highlight the interplay between excitonic condensation and other collective phenomena, such as charge density waves and spin-orbit coupling, providing a rich platform for exploring correlated electronic states.\cite{Kogar2017, Varsano2020} These insights are underpinned by first-principles many-body calculations, including the GW-Bethe-Salpeter equation framework, which captures the delicate balance between Coulomb interactions and band structure effects.\cite{GW_1,GW_2,GW_3,GW_4,GW_Graphene,GW_NT_1,GW_NT_2,GW_Graphene,Book_GW,Steinhoff2014} The EI phase offers exciting prospects for applications in optoelectronics and quantum information, particularly in strongly correlated two-dimensional materials.

Several theoretical studies, including those by Gao et al.\cite{Gao2023} and Grytsiuk et al.,\cite{Grytsiuk2024} have described Nb$_3$Cl$_8$ as an electronic Mott insulator using single- and multi-orbital Hubbard models, while neglecting excitonic effects, to explain features such as the optical absorption peak at 1.12 eV.\cite{Sun2022} These studies assume a metallic phase with half-filled bands at the Fermi level, leading to the emergence of lower and upper Hubbard bands through strong correlation effects.

\textcolor{black}{Our first-principle GW-BSE calculations for bulk Nb$_3$Cl$_8$ yield a bright exciton (SI) consistent with experiment,\cite{liu2024possible} compatible with a single-band Mott state hosting $S = \tfrac{1}{2}$ moments per Nb$_3$ trimer.\cite{Gao2023, Grytsiuk2024} Using GW$-$BSE calculations for bulk Nb$_3$Cl$_8$, the first absorption peak appears at 1.15~eV, which is in excellent agreement with the experimental values.\cite{Sun2022} In contrast to the bulk material, our GW-BSE calculations for SL Nb$_3$Cl$_8$, owing to the reduced screening in the SL limit,  indicate the presence of a dark exciton ground state at a negative energy $E_{\rm EMI} = -0.14$ eV, i.e. being 0.14 eV below the GW-renormalized band gap, with a binding energy of 2.64 eV, providing compelling evidence for an excitonic insulator (EI) phase based on a dark spin-1 triplet exciton. To distinguish from exciton condensates,\cite{Zhu&Rice1995} we propose that SL Nb$_3$Cl$_8$ exhibits a novel EI insulator phase, namely an Excitonic Mott Insulator (EMI) phase. Regarding the periodicity of the electrostatic charge distribution, our calculations reveal that each Nb trimer hosts an  out-of-plane electric dipole moment with strong electric anisotropy energy, exceeding the inter-exciton electric dipole–dipole interaction by two orders of magnitude. Therefore, the electric dipole moments of all the localized Frenkel excitons remain pinned in a single out-of-plane direction, thereby keeping the periodicity of the primitive unit cell. }

\textcolor{black}{Considering our GW-BSE results for bulk and SL Nb$_3$Cl$_8$, we propose a screening/thickness-driven crossover: bulk-like samples realize the electronic $S = \tfrac{1}{2}$ Mott/QSL regime reported experimentally,\cite{liu2024possible} compatible with Grytsiuk et al.'s theory,\cite{Grytsiuk2024} keeping in mind that electron tunneling between the layers is almost absent, while the true monolayer may approach a spin-1 exciton--Mott limit, as shown schematically in Fig.~\ref{fig:elec_exc_hb_model}. We propose dark electron spin resonance (ESR) or electron paramagnetic resonance (EPR) measurements on monolayer and thicker ($\geq$10~nm) Nb$_3$Cl$_8$ flakes. A zero-field splitting (ZFS) in the resonance spectrum would indicate integer-spin ($S=1$) excitons, whereas a single $g \!\approx\! 2$ line without ZFS would confirm half-integer spins ($S=\tfrac{1}{2}$).}

Using this state as an intermediate point, we then use a mean-field spin-1 Bose-Hubbard model with a three times larger unit cell, allowing for a variety of magnetic orderings, to determine the ground state also with respect to the spin degree of freedom of the dark spin-1 Frenkel excitons, revealing a 120$^\circ$ in-plane spin-1 configuration.
Our study provides insight into the electronic structure and flat band-induced excitonic structure in SL Nb$_3$Cl$_8$ with a breathing Kagome lattice structure, thereby offering a valuable framework for exploring the interaction between geometry and electron correlations.

\section{Results}
\subsection{Numerical Results}

\begin{figure*}[!t]\includegraphics[width=\textwidth]{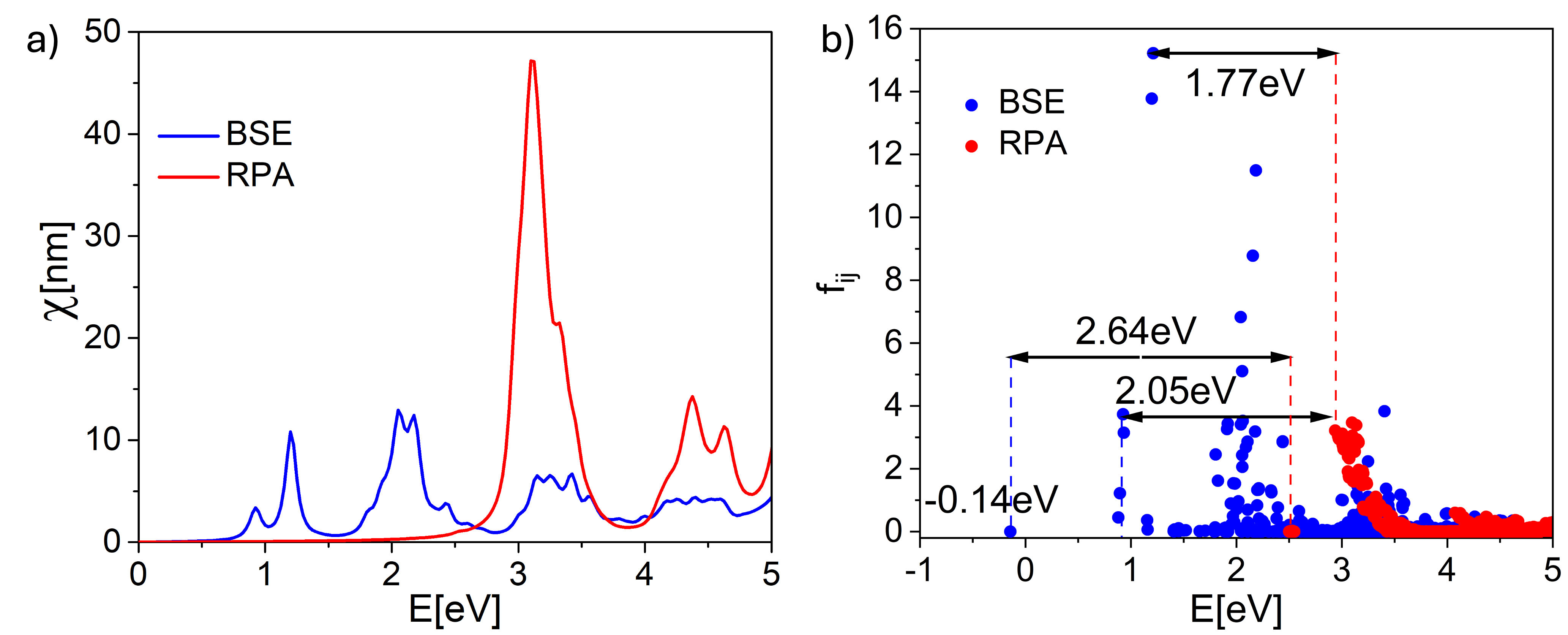}
\centering
\caption{a) The absorption spectra of SL Nb$_3$Cl$_8$, shown without (dashed black curve) and with (solid blue curve) electron-hole interactions, exhibit a prominent peak at 1.2 eV. This peak aligns remarkably well with experimental optical absorption measurements, highlighting the precision of the GW calculations. b) The oscillator strength f$_{ij}$ as a function of energy reveals key insights into the system's excitonic properties. The Bethe-Salpeter Equation (BSE) solution at E$_{\rm EMI}=-0.14$ eV, which corresponds to the IPA state at 2.5 eV, giving a binding energy of 2.64 eV, E$_{ev}<0$ exhibits clear signatures of an exciton insulator. In contrast, the first two bright exciton peaks can be seen at 0.93 eV and 1.2 eV with binding energies 2.05 eV and 1.77 eV, respectively.}
\label{fig:susceptibility_TD}
\end{figure*}
 
\paragraph{Electronic Properties:}Our model system consists of a single layer (SL) of Nb$_3$Cl$_8$ with 11 atoms in the unit cell, where a layer of Nb atoms is sandwiched between layers of Cl atoms, held together through strong covalent bonds as shown in Fig.~\ref{fig:Nb3Cl8_Structure_Bs} (a). A breathing Kagome lattice is formed by Nb atoms creating irregular hexagons with two sets of sides of different lengths, resulting in two sets of outer equilateral triangles with different areas, as shown in Fig.~\ref{fig:Nb3Cl8_Structure_Bs} (a). The electronic band structure of Nb$_3$Cl$_8$ is largely determined by the symmetry properties of the high-symmetry points and lines in the first Brillouin zone of its triangular lattice.  Fig.~\ref{fig:Nb3Cl8_Structure_Bs} also illustrates both the unit cell and the Brillouin zone of Nb$_3$Cl$_8$. The crystal structure of single-layer (SL) Nb$_3$Cl$_8$ exhibits C$_{3v}$ symmetry.

Fig.~\ref{fig:Nb3Cl8_Structure_Bs} (b) presents the band structure results from noncollinear SOC calculations, incorporating self-energy corrections through GW calculations. Nearly flat bands with a very small dispersions can be seen, well separated from the continuum of bands and are therefore available for direct experimental observation.  A substantial increase of 2 eV in the band gap with the GW correction compared to the PBE (SI) highlights the importance of self-energy corrections in Nb$_3$Cl$_8$. In the Supplementary Material, we provide compelling evidence from both non-spin-polarized (NSP) and spin-polarized (SP) calculations for the magnetic nature of SL Nb$_3$Cl$_8$. In NSP calculations, the electronic states at the Fermi level are degenerate. However, in spin-polarized (SP) DFT calculations, these states split into valence and conduction bands with opposite spin orientations. As a result, optical transitions from the valence to the conduction band are forbidden due to spin selection rules, leading to the formation of a dark exciton. Interestingly, NSP DFT results indicate that SL Nb$_3$Cl$_8$ behaves as a metallic system,\cite{grytsiuk2024nb3cl8,PhysRevLett.130.136003} highlighting the critical role of spin polarization in accurately capturing the material's magnetic and optical properties. Since we obtain that the $120^\circ$ spin configuration has the lowest energy, we choose to perform non-collinear DFT calculations. It is interesting to note that flat bands appear in the form of triplets, i.e. a singlet accompanied by a two-fold degenerate doublet. The existence of triplets is a consequence of three-fold rotational symmetry of the crystal.\cite{Erementchouk_MoS2,Khan_TMDCs} 

Since we are interested in the FBs originating from the BKL formed by Nb atoms, we project the FBs in the band structure, onto the d-orbitals of the Nb atoms to investigate the orbital character of the FBs (SI). It can be seen that substantial d$_{z^2}$ orbital contribution is present in conduction and valence bands. In the analytical modeling section, we develop a single symmetric orbital tight-binding model that corroborates the numerical results, offering further validation and insight into the system's behavior. \textcolor{black}{In particular, we show here that a trimer of Nb atoms contains an electric dipole moment with a very large out-of-plane anisotropy energy, which is two orders of magnitude larger than the electric dipole-dipole interaction. Therefore the primitive unit cell of Nb$_3$Cl$_8$ captures fully the periodicity of the electric dipole moment for numerical calculations.}

\paragraph{Optical Properties:} Fig.~\ref{fig:susceptibility_TD} presents the absorption spectra with oscillator strengths for SL Nb$_3$Cl$_8$, both with and without electron-hole interactions. From the oscillator strength it is evident that the first exciton eigenvalue at $E_{\rm EMI}=-0.142$ eV is indeed dark. A negative BSE eigenvalue indicates that the ground state exciton possesses an exceptionally large binding energy of 2.64 eV, greater than the material's bandgap. This phenomenon is known as the EI phase,\cite{PhysRevLett.126.196403} where excitonic interactions dominate over free charge carriers. Notably, the BSE ground state corresponds to a dark exciton which couples inefficiently with light. Dark excitons can be populated when a material is cooled, as excitons in higher-energy bright states lose energy and transition to lower-energy dark exciton states. Since the Frenkel excitons are strongly localized on the Nb atoms, we propose a new EI phase for Nb$_3$Cl$_8$ called EMI phase.


Although the low-dimensional exciton insulator phase has been proposed in numerous theoretical studies,\cite{PhysRevLett.126.196403,spin_triplet_SHGr,liu2024one} experimental realization of such systems \cite{PhysRevLett.126.196403,spin_triplet_SHGr,liu2024one} remains a formidable challenge, demanding precise atomic-scale engineering and control; in contrast, Nb$_3$Cl$_8$ has already been successfully realized, marking a significant step forward in this field. 

The first two bright excitons appear at energies of 0.94~eV and 1.2~eV, with binding energies of 2.64~eV and 2.05~eV, respectively, as shown in Fig.~\ref{fig:susceptibility_TD}. These unusually high binding energies reflect the weak dielectric screening in SL Nb$_3$Cl$_8$, a characteristic feature of many two-dimensional materials.\cite{MoS2_EBE,Luo_2025} Such strong excitonic effects highlight the potential of SL Nb$_3$Cl$_8$ for optoelectronic applications, particularly in devices that rely on strong light–matter interactions.

We also calculate the optical spectrum of bulk Nb$_3$Cl$_8$ using the GW$+$BSE approach and find a band gap of 1.15~eV (SI), which is in excellent agreement with the experimentally observed value of 1.12~eV.\cite{Sun2022} This close alignment with experimental data underscores the precision and reliability of the GW and BSE methods in accurately capturing the optical properties of the system.

Due to the very weak interlayer coupling\cite{Haraguchi2024} in bulk Nb$_3$Cl$_8$, it is meaningful to draw a direct comparison between the dark and the first two bright exciton states in single-layer (SL) Nb$_3$Cl$_8$ and their counterparts in the bulk. A detailed comparison of the optical properties of bulk and SL Nb$_3$Cl$_8$ is presented in Table~\ref{tab:Comparison_SL_Bulk}. The exciton binding energies E$_b$ in bulk Nb$_3$Cl$_8$ remain substantially high, highlighting the significant role of flat electronic bands in enhancing excitonic effects.
\begin{table}[]
    \centering
    \begin{tabular}{|c|c|c|c|}
    \hline
         \multicolumn{2}{|c|}{Optical Properties} & E$_g$ & E$_b$  \\
         \hline
         \hline
         SL Nb$_3$Cl$_8$ & \makecell[l]{Dark Exciton \\ 1st Bright Exciton \\ 2nd Bright Exciton} & \makecell[l]{$-$0.14~eV \\ 0.94~eV \\ 1.21~eV} & \makecell[l]{2.64~eV \\ 2.05~eV \\ 1.77~eV} \\
         \hline
         Bulk Nb$_3$Cl$_8$ & \makecell[l]{Dark Exciton \\ 1st Bright Exciton \\ 2nd Bright Exciton} & \makecell[l]{0.16~eV \\ 1.15~eV \\ 1.51~eV} & \makecell[l]{1.77~eV \\ 1.16~eV \\ 0.81~eV}\\
         \hline
    \end{tabular}
    \caption{Comparison of band gap E$_g$, exciton binding energies E$_b$ of bulk and SL Nb$_3$Cl$_8$.}
    \label{tab:Comparison_SL_Bulk}
\end{table}

\paragraph{Optical Selection Rules:}Nb$_3$Cl$_8$ possesses hexagonal symmetry, characterized by a three-fold rotational axis and broken inversion symmetry. Much like a hydrogen atom, an exciton confined to a two-dimensional plane exhibits excitonic angular momentum (EAM) originating from the orbital motion of the electron relative to the hole. Despite inversion-symmetry breaking, we focus on the $\Gamma$-point to bypass the need to account for the conservation of valley angular momentum (VAM), which typically arises from broken inversion asymmetry. Additionally, the three-fold rotational symmetry transforms the incoming angular momentum into a modulus of three by absorbing the excess angular momentum into the lattice. Consequently, under normally incident light, the conservation of out-of-plane angular momentum leads to the following optical selection rule:
\begin{equation}
    \Delta m\hbar=\Delta l_{ex}\hbar+3N\hbar
    \label{Eq:selection_rule}
\end{equation}
\begin{figure}[b]\includegraphics[width=1.5in]{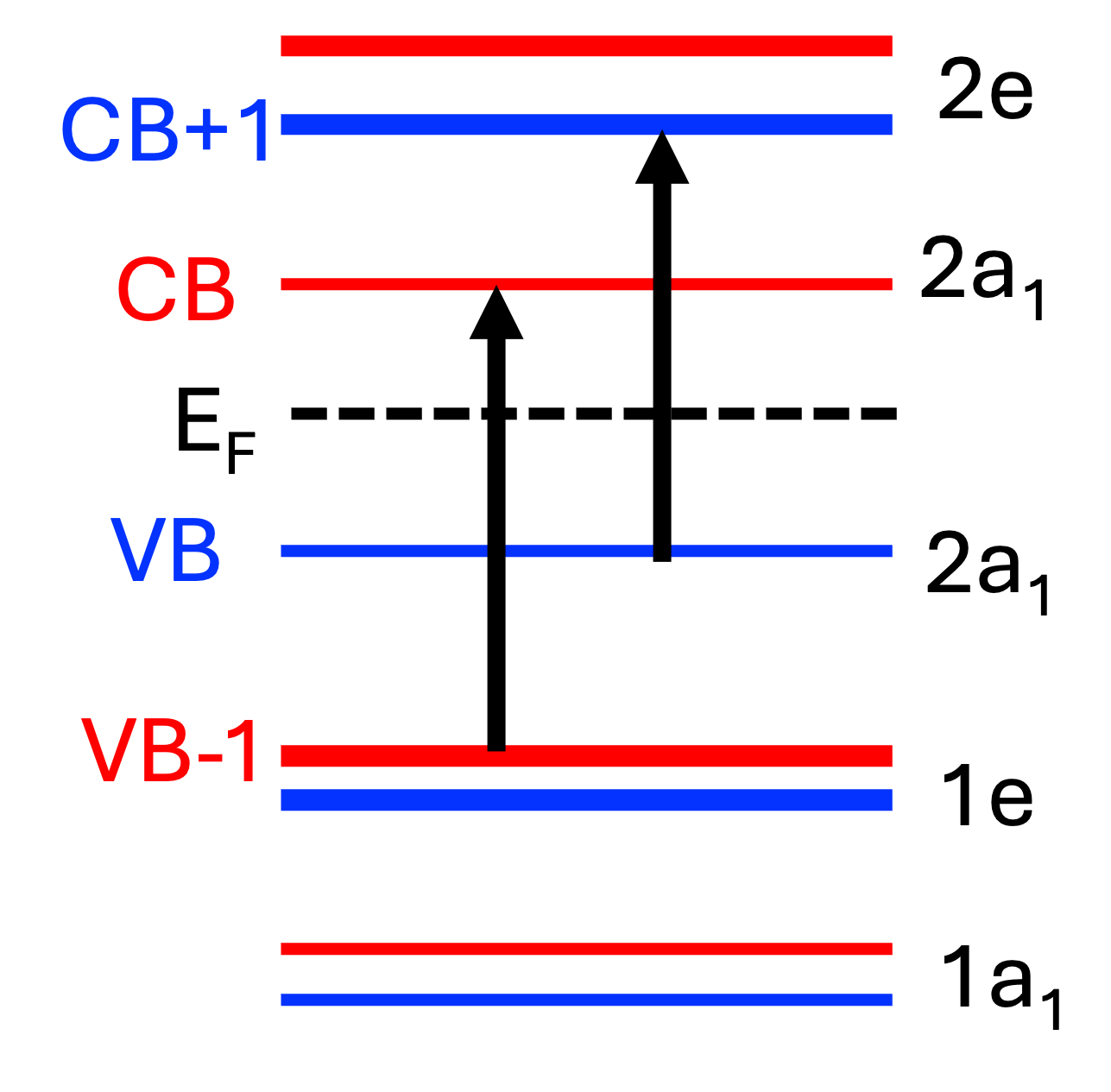}
\caption{Molecular orbital configuration constructed in the Nb$_3$ trimer in Nb$_3$Cl$_8$. Blue (red) line show spin-up (spin-down) states with no preferential direction. Optical transition between VB and CB is forbidden because of the spin selection rules, which makes the ground state spin-triplet exciton dark. Black arrows show two of the optically allowed transitions between 2a$_1$ (VB) and 2e (CB+1), corresponding to the bright exciton at 0.93 eV, and between 1e (VB-1) and 2a$_1$ (CB), corresponding to the bright exciton at 1.2 eV (see Fig.~\ref{fig:BSE_fatbands}).}
\label{fig:Schematic_Energy_level}
\end{figure}
Upon photon absorption, the change in momentum, $\Delta m\hbar$, of the incident photon induces a corresponding change in the angular momentum of the exciton, $\Delta l_{ex}\hbar$, and the lattice, quantified as $3N\hbar$, where $N=0, \pm1, \pm2,\dots$. Eq.~\eqref{Eq:selection_rule} illustrates the selection rules governing transitions between the d$_{z^2}$ and \{d$_{x^2-y^2}$,d$_{xy}$\} orbitals, which require $N=\pm3$. In contrast, transitions from the orbitals d$_{z^2}$ to d$_{yz}$ and d$_{xz}$ occur when $N=0$. Notably, out-of-plane transitions between d$_{z^2}$ orbitals are also allowed for $N=0$. This entire description can be rigorously explained using a group-theoretical analysis of the C$_{3v}$ symmetry. Within the electric dipole approximation, transitions between orbitals belonging to different irreducible representations (IRs) are summarized in Table SI in the Supplementary Information. In the electric dipole approximation the optical transitions are governed by the matrix element $\langle \psi_i|x_j|\psi_f\rangle$. The initial state $\psi_i$, the final state $\psi_f$, and the position operator $x_j$ transform according to the IRs $\Gamma(\psi_i)$, $\Gamma(\psi_f)$, and $\Gamma(x_j)$, respectively. An electric dipole transition between two states is allowed if the direct product $\Gamma(\psi_i) \otimes \Gamma(x_j) \otimes \Gamma(\psi_f)$ includes $\Gamma(I)$ in its decomposition into a direct sum. Group theoretical description corroborates the selection rules depicted in Eq.~\eqref{Eq:selection_rule}. It is important to emphasize that the electronic transitions are localized in real space, where the symmetry at the center of the trimer is governed by the C$_{3v}$ point group. Accordingly, we utilize the C$_{3v}$ symmetry operations and their corresponding irreducible representations to derive and explain the optical selection rules, which are shown in Table \ref{tab:c3v_symmetry}.

\begin{table}[t!]
    \centering
    \setlength{\tabcolsep}{5pt}
    \begin{tabular}{|c|c|c|c|}
    \hline
         C$_{3v}$ & E & 2C$_3$(z) & 3$\sigma_v$ \\
         \hline
         \hline
         A$_1$ & +1 & +1 & +1 \\
         \hline
         A$_2$ & +1 & +1 & -1\\
         \hline
         E & +2 & -1 & 0 \\
         \hline
         
    \end{tabular}
    \setlength{\tabcolsep}{5pt}
    \begin{tabular}{|c|c|c|c|}
    \hline
         C$_{3v}$ & A$_1$ & A$_2$ & E \\
         \hline
         \hline
         A$_1$ & $\pi$ &  & $\sigma$ \\
         \hline
         A$_2$ &  & $\pi$ & $\sigma$\\
         \hline
         E & $\sigma$ &  & $\pi$ \\
         \hline
         
    \end{tabular}
    \caption{Left: character table of C$_{3v}$ symmetry. Right: Allowed dipole transitions. $\sigma$ and $\pi$ represents in-plane and out-of-plane transitions.}
    \label{tab:c3v_symmetry}
\end{table}

Fig.~\ref{fig:Schematic_Energy_level} presents a schematic diagram of the molecular orbital configuration along with the corresponding irreducible representations of a Nb$_3$ trimer in Nb$_3$Cl$_8$.
Although out-of-plane transitions between the conduction and valence bands are allowed based on orbital character, these transitions are forbidden due to spin selection rules. Therefore, the spin-triplet Frenkel exciton based on the CB and VB Bloch states, corresponding to the 2a$_1$ molecular orbital, is dark. There are two symmetry-allowed optical transitions between the 2a$_1$ spin-up state (VB) and the 2e spin-up state (CB+1), corresponding to the bright exciton at 0.93 eV, and between the 1e spin-down state (VB-1) and the 2a$_1$ spin-down state (CB), corresponding to the bright exciton at 1.2 eV (see Fig.~\ref{fig:BSE_fatbands}).

\begin{figure*}[!t]
\centering
\includegraphics[width=\textwidth]{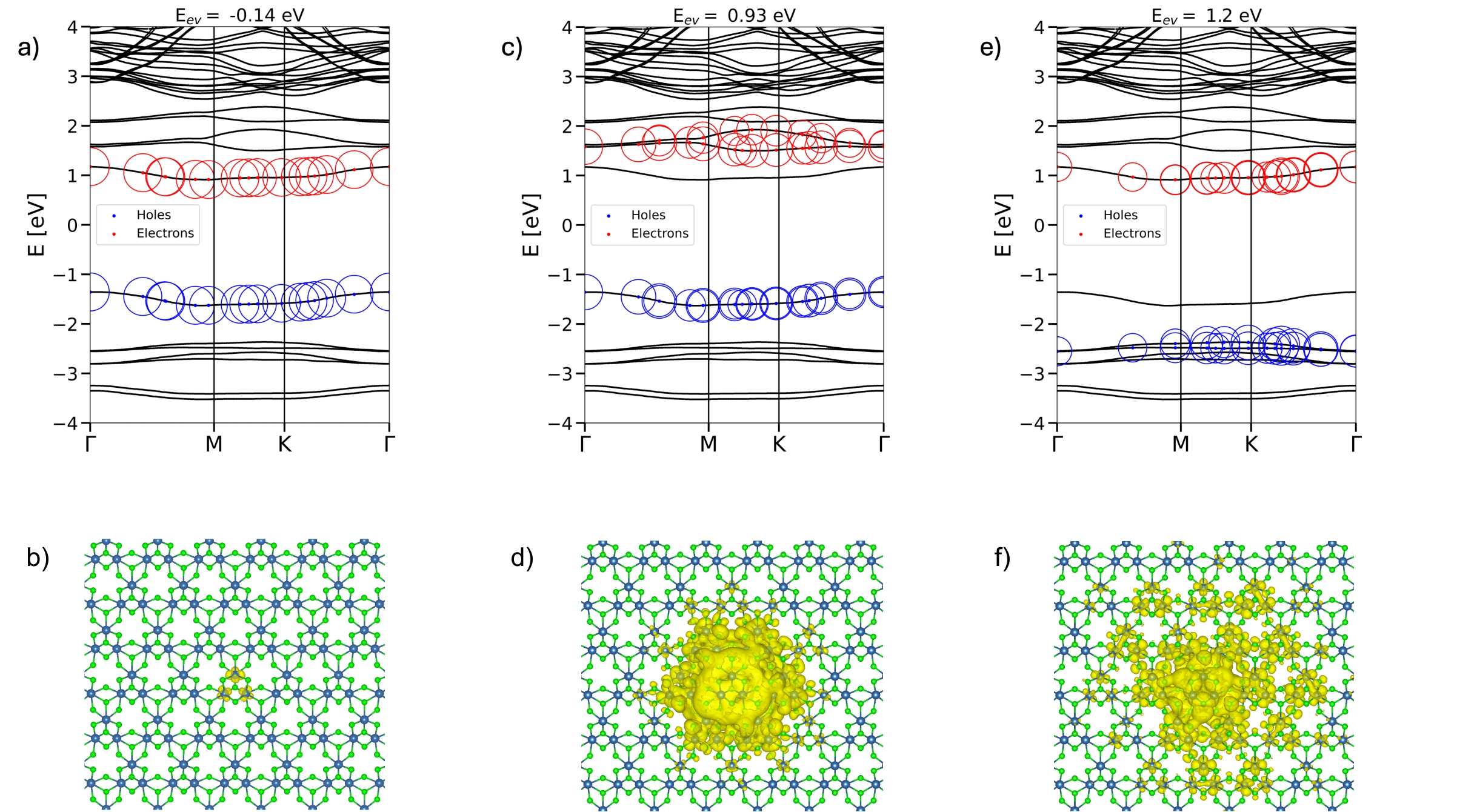}
\caption{a) Fat bands obtained for the BSE ground state, corresponding to the eigenenergy of -0.14 eV (dark Frenkel exciton), are mapped onto the electronic band structure, helping to identify which electron-hole Bloch states contribute most significantly to the BSE eigenstate. Radius of the red (blue) circles show the amplitude in the exciton wavefunction. b) The exciton wavefunction distribution in real space for the dark Frenkel exciton ground state.  c) Fat bands are mapped onto the electronic band structure for the bright exciton with eigenenergy 0.93 eV. d) The exciton wavefunction for the bright exciton at 0.93 eV. e) Fat bands are mapped onto the electronic band structure for the bright exciton with eigenenergy 1.2 eV. f) The exciton wavefunction for the bright exciton at 1.2 eV.}
\label{fig:BSE_fatbands}
\end{figure*}
\begin{figure}[t]\includegraphics[width=3.2in]{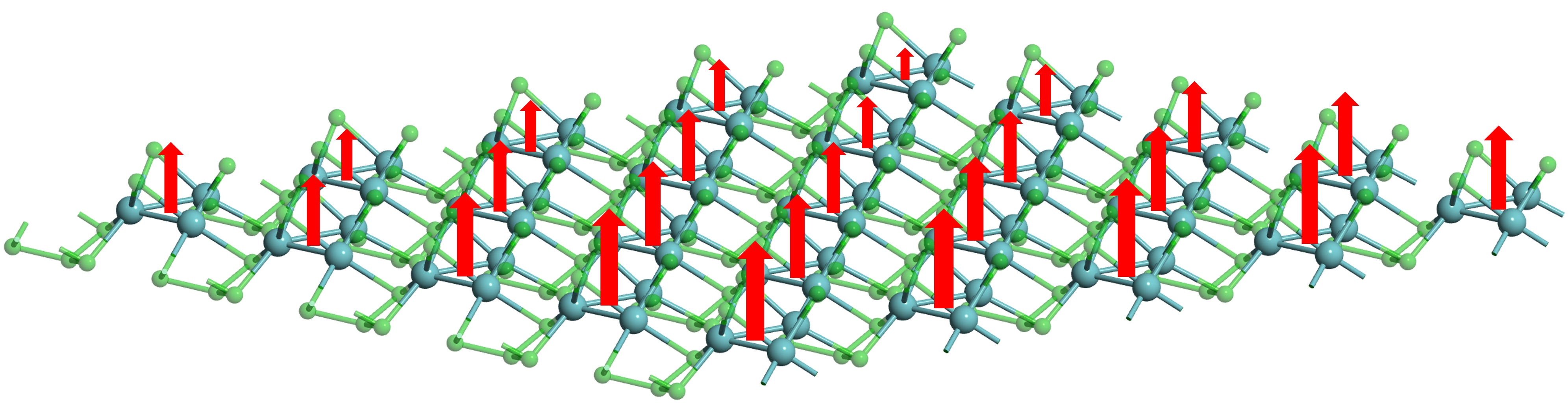}
\centering
\caption{There is one Frenkel exciton on each trimer of Nb atoms (dark green balls). Each Frenkel exciton carries an electric dipole (red arrows). The electric dipoles are oriented perpendicular to the plane of SL Nb$_3$Cl$_8$.}
\label{fig:ED}
\end{figure}


\subsection{The Dark Exciton Ground State}

Solving the two-band BSE for Nb$_3$Cl$_8$, we find that the lowest energy excitonic state is a dark exciton, with an energy of $E_{\rm EMI}=-0.14$ eV with a binding energy of 2.64 eV. This dark exciton arises from the fact that the conduction and valence bands have opposite spin, forming a spin-triplet exciton state, which effectively doubles the magnetic dipole moment on a trimer from $s=1/2$ to $S=1$ when compared with the electronic ground state obtained by DFT only. The exciton wavefunction is centered near the $\Gamma$-point in momentum space. The delocalization in $\bk$-space (Fig.~\ref{fig:BSE_fatbands} (a and b)) indicates a small exciton radius in real space. To estimate its Bohr radius, we calculate the effective masses of the electron and hole in the CB and VB at the $\Gamma$-point. We obtain $m_e^*=2.444m_e$ for the electron and $m_h^*=2.559m_e$, resulting in an effective reduced mass of $\mu=1.25m_e$. Thus, for the Bohr radius we obtain $R_B=1.74$ \AA, (for $\epsilon_{\parallel}=1.818$), indicating localization on a Nb site. Note that the Bohr radius is close to the radius of the Nb atom, which is $R_{\rm Nb}=1.46$ \AA. Thus, we identify this lowest-energy dark exciton being a Frenkel exciton. In addition we also show in SI (Fig.~S5), the real spce plot of the dark exciton state, which is localized on the trimer with a radius of 1.46 \AA, which is quite close to the Bohr radius R$_H$.
\begin{figure*}[]
\includegraphics[width=\textwidth]{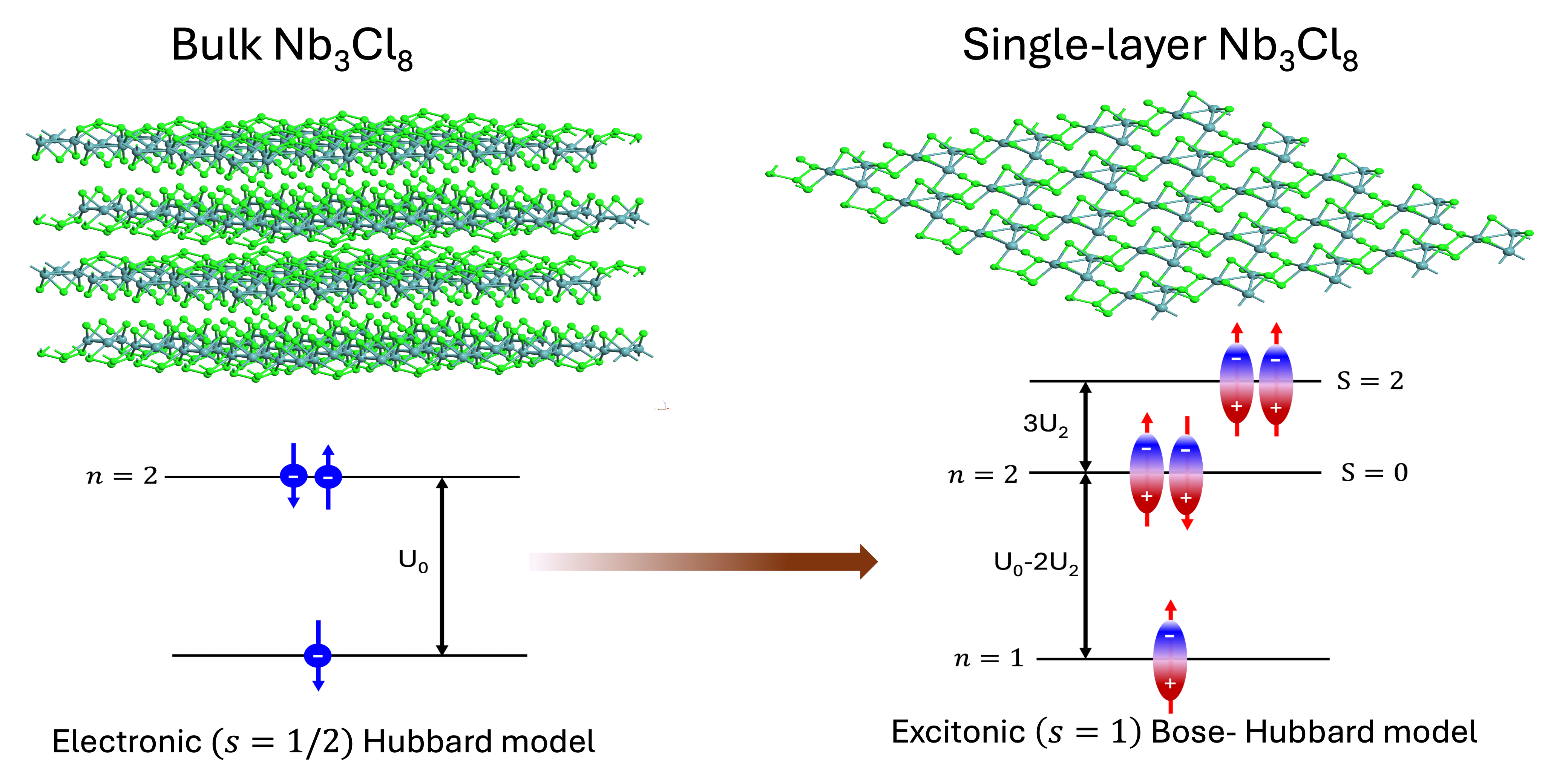}
\caption{Schematic evolution of correlation physics in Nb$_3$Cl$_8$ from bulk to SL. In the bulk phase (left), electronic screening leads to electron–electron correlations described by the standard Hubbard model. In the SL limit (right), screening is reduced, stabilizing spin-1 excitons that behave as interacting bosons governed by a Bose–Hubbard framework.}
\label{fig:elec_exc_hb_model}
\end{figure*}

Furthermore, our theoretical analysis of the tight-binding Hamiltonian for a spin-1 particle on a trimer supports these findings. Specifically, we identify the Frenkel exciton to be in the 2a$_1$ ground state with orbital angular momentum \( m_L = 0 \). This state arises from the symmetric combination of the \( d_{z^2} \) orbitals of the Nb atoms in the trimer, leading to a localized Frenkel exciton with total spin \( S = 1 \).

\subsection{Electric Dipole Ordering based on Spin-Triplet Frenkel Exciton}
While in the absence of SOC, spin-triplet excitons are optically inactive (dark) due to spin selection rules, they exhibit an electric dipole moment due to their $p$-like orbital wave function. Spin triplet excitons are characterized by their total spin quantum number \( S = 1 \), which arises from the parallel alignment of the electron and hole spins. Since electron and hole are fermions, if the spin part of the exciton is symmetric, the orbital part must be antisymmetric, e.g. $p$-like.

The spin triplet state comprises three possible spin projections, corresponding to the eigenvalues of the total spin projection operator \( S_z \), i.e.
\( |S=1, m_S = +1\rangle = |\uparrow \Uparrow\rangle \),
    \( |S=1, m_S = 0\rangle = \frac{1}{\sqrt{2}} \left( |\uparrow \Downarrow\rangle + |\downarrow \Uparrow\rangle \right) \), and
   \( |S=1, m_S = -1\rangle = |\downarrow \Downarrow\rangle \).

The antisymmetric nature of the spatial wavefunction means the lowest-energy orbital state is \( p \)-like, with an angular momentum quantum number \( \ell = 1 \) and magnetic quantum number $m_L=-1,0,+1$. Thus, the Frenkel excitons can carry an electric dipole moment due to the separation of the electron and hole. 
The electric dipole-dipole interaction between two electric dipoles \( \bp_i \) and \( \bp_j \), located at positions \( \br_i \) and \( \br_j \), is given by
\begin{equation}
 H_{\text{dd}} = \frac{1}{4\pi\epsilon_0\epsilon_r} \frac{1}{2}\sum_{ij,j\ne i} \left[ \frac{\bp_i \cdot \bp_j}{r_{ij}^3} - 3\frac{(\bp_i \cdot \br_{ij})(\bp_j \cdot \br_{ij})}{r_{ij}^5} \right],  
 \label{Eq:Dipole_Dipole_Interaction}
\end{equation}

\noindent where \( \br_{ij} = \br_j - \br_i \) is the displacement vector between the two electric dipoles and \( r_{ij} = |\br_{ij}| \) is the distance between them. $\epsilon_r$ is the relative permittivity of the material. In the case of Nb$_3$Cl$_8$ the dielectric constant exhibits uniaxial anisotropy with the in-plane value estimated by BSE of $\epsilon_\parallel=1.818$ and out-of-plane value of $\epsilon_\perp=1.1285$.

For a triangular lattice, the geometry dictates the relative orientations of the dipoles. 
We calculated the electric dipole moment of dark exciton using exciton wave function 
\begin{equation}
\mathbf{d} = \int \mathbf{r}\Psi^*_\lambda\left(\mathbf{r}\right)\Psi_\lambda\left(\mathbf{r}\right)d\mathbf{r}    
\end{equation}
\noindent where $\Psi_\lambda(\mathbf{r}) = \Psi_\lambda(\mathbf{r}_e-\mathbf{r}_h)$ and position of the hole corresponds to the center of the Nb$_3$ triangle. The value of the dipole moment is $d=0.733$ Debye, and the orientation is along $z$-direction. We calculated the energy of interaction of such electric dipole moments placed at each Nb$_3$ triangles for different configurations using Eq.~(\ref{Eq:Dipole_Dipole_Interaction}).
We found that ferroelectric in-plane orientation gives $-0.0087$ eV per Nb$_3$ triangle, whereas the energy for the out-of-plane ferroelectric configuration is $0.0171$ eV per Nb$_3$. Thus the reorientation of such dipole moments of excitons from in-plane to out-of-plane requires $0.0258$ eV. However, if to calculate the interaction of such dipole moments with crystal field of Nb$_3$Cl$_8$ as
\[
E_{\mathbf{dE}} =-\int \mathbf{r}\Psi^*_\lambda\left(\mathbf{r}\right)\Psi_\lambda\left(\mathbf{r}\right)\mathbf{E}_{cr}\left(\mathbf{r}\right)d\mathbf{r}
\]
we found that the energy of interaction of dipoles in ferroelectric configuration with crystal field is $-7.08$ eV per Nb$_3$  triangle, which is much larger that the interaction between dipoles, because the charge distribution for exciton is mostly around Nb ions, where the crystal field is the strongest. To calculate crystal field, we used the point charge model and for charges associated with the ions, we used Bader charge analysis, which gives charge on Cl atom $-0.6$ and on Nb ion $+1.6$. Thus, the interaction of electric dipole moments with the crystal field is much stronger, than the interaction of dipole moments between each other. The preferable orientation of excitonic dipoles is in $z$-direction making ferroelectric configuration.  

\subsection{Bose-Hubbard Model}
We use the results from the DFT-GW-BSE calculations as the starting point for Bose-Hubbard mean field calculations. Following Ref.~\onlinecite{Tsuchiya_BHM} we use the Bose-Hubbard model for spin-1 excitons (on trimers), which describes interacting bosons on a triangular lattice (of trimers) with 6 nearest neighbors, incorporating both spin-dependent and density-dependent interactions. The Hamiltonian for this system can be written as
\begin{equation*}
\begin{split}   
    \hat{H}=-t\sum_{\langle i,j\rangle}\sum_{\alpha}\left(\hat{a}^\dagger_{i\alpha}\hat{a}_{j\alpha}+c.c\right)-
    \sum_{i}\sum_{\alpha}\left(\mu-\epsilon_{i}\right)\hat{a}^\dagger_{i\alpha}\hat{a}_{i\alpha}
    \\
    +\frac{1}{2}U_0\sum_{i}\sum_{\alpha,\beta}\hat{a}^\dagger_{i\alpha}\hat{a}^\dagger_{i\beta}\hat{a}_{i\beta}\hat{a}_{i\alpha}
    \\
    +\frac{1}{2}U_2\sum_{i}\sum_{\alpha,\beta,\delta,\gamma}\hat{a}^\dagger_{i\alpha}\hat{a}^\dagger_{i\gamma}\textbf{S}_{\alpha\beta}\cdot\textbf{S}_{\gamma\delta} \hat{a}_{i\delta}\hat{a}_{i\beta}.
\end{split}
\end{equation*}
The first term on the right is the hopping term with hopping probability $t$ between nearest neighbors. $\langle i,j\rangle$ expresses a summation over nearest-neighbor sites. $\hat{a}_{i\alpha}$ and $\hat{a}^{\dagger}_{i\alpha}$ are the annihilation and creation operators for an exciton at site $i$ with spin components $\alpha={1,0,-1}$. $\epsilon_i$ is the onsite energy. $\mu$ is the chemical potential, which controls the number of excitons on a lattice site. $U_0$ is the onsite spin-independent Coulomb interaction while $U_2$ is the onsite spin-dependent Coulomb interaction. These two types of Coulomb interaction are defined by
\begin{equation*}
    U_F=c_F\int d\textbf{r}|w_0(\bf{r}-\bf{r}_i)|^4,
\end{equation*}
where $F=0,2$, and
$w_0$ is the Frankel exciton wavefunction at a site. We consider low-energy $s$-wave scattering, for which the total spin states $S = 0$ and $S = 2$ are allowed due to the symmetry of the wavefunction, with $c_F$  being the scattering lengths for two colliding excitons at a site.
The model is solved for $n_i = 1$ (one exciton per lattice site) and $S_i = 1$ (spin-1 excitons). First, the model is solved exactly in the strongly correlated limit ($t = 0$). The hopping term is then introduced perturbatively using a mean-field decoupling approach. The resulting phase diagram is shown in Fig.~\ref{fig:BHM}. The yellow square in the phase diagram indicates the ground state of the exciton insulator phase for trimers forming a triangular lattice with hopping $t=0.022$ eV and $\mu=0.4$ eV, located at the center of the Mott gap. Thus, we conclude that   the exciton insulator phase in Nb$_3$Cl$_8$ is located well within the Mott insulator region.

In stark contrast to the conventional electronic Hubbard model\cite{Grytsiuk2024}, we propose a spin-1 exciton Bose–Hubbard model to demonstrate the emergence of an EMI phase in SL Nb$_3$Cl$_8$. 

\subsection{Magnetic Ordering based on Spin-Triplet Frenkel Exciton}
\label{sec:magnetic_ordering}
To derive an effective Bose-Hubbard Hamiltonian for the spin-1 Frenkel exciton, we note that its binding energy of $E_b=2.64$ eV is much larger than the intra-trimer hopping energy $t_0=-0.325$ eV\cite{Grytsiuk2024}, which is much larger than the inter-trimer hopping energy $t_1=0.022$ eV.\cite{Grytsiuk2024}
Since the spin system is based on the single $d_{z^2}$ Nb orbital, for which $m_l=0$, and, in addition, the molecular orbital 2a$_1$ also has $m_L=0$,  we infer that SOC can be neglected, in agreement with previous theoretical studies.\cite{Grytsiuk2024} 

\begin{figure}[!t]\includegraphics[width=3.3in]{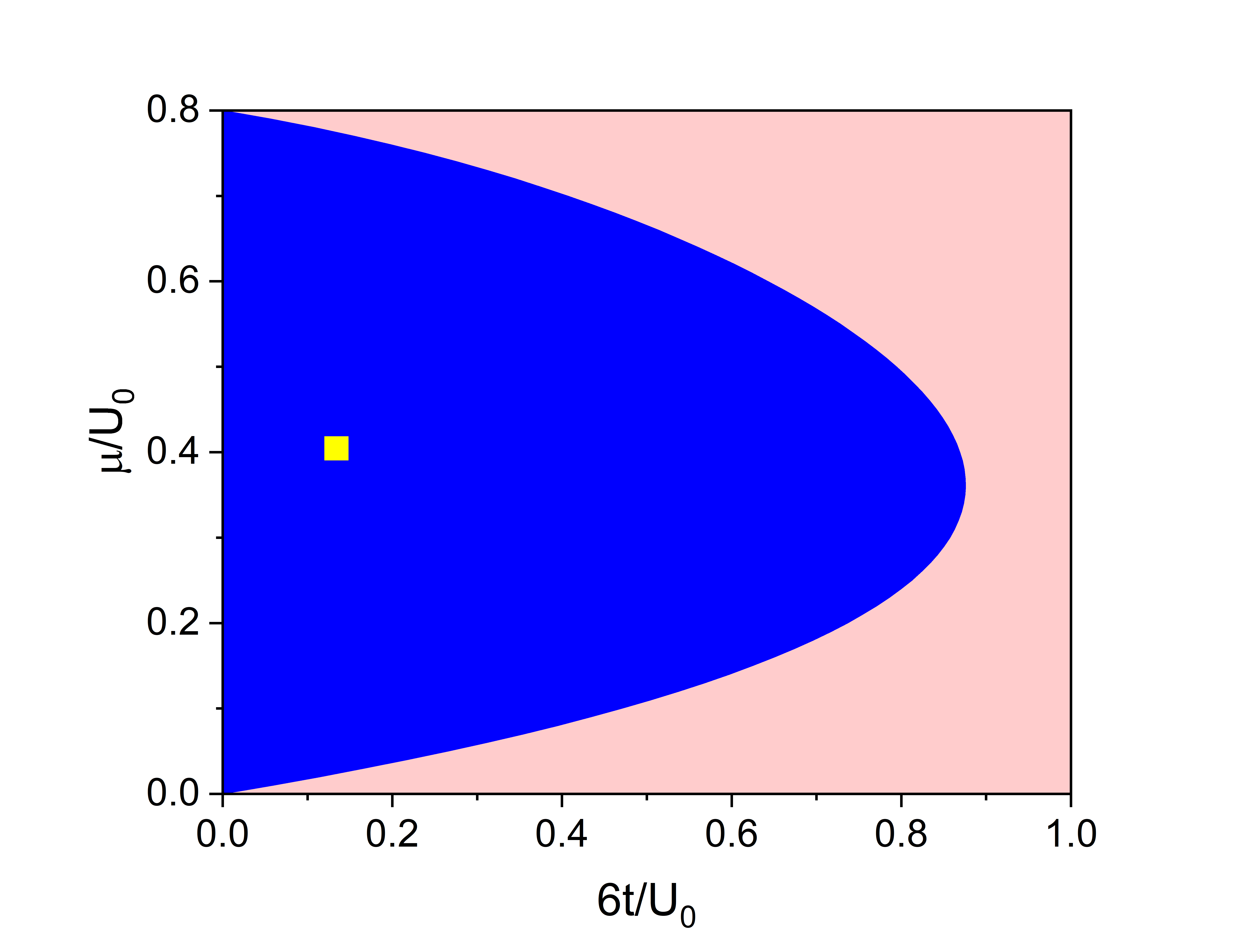}
\caption{ Phase diagram of the Bose-Hubbard model for spin-1 ($S_i = 1$) excitons on a triangular lattice with six nearest neighbors, shown for an interaction ratio $U_2/U_0 = 0.1$ ($U_0=$1 eV). The blue region represents the Mott insulating phase, while the pink region corresponds to the superfluid phase. The yellow dot marks the point $6t = 0.132$ eV and $\mu = 0.4$ eV.}
\label{fig:BHM}
\end{figure}
Therefore, we obtain the effective exchange Hamiltonian
\begin{equation}
H_{\text{eff}} = J \sum_{\langle i,j \rangle} \mathbf{S}_i \cdot \mathbf{S}_j,
\end{equation}
where \( J \) is the exchange coupling between neighboring trimers, \( \mathbf{S}_i \) is the spin-1 operator on trimer \( i \), and \( \langle i,j \rangle \) denotes neighboring trimers.
The exchange coupling \( J \) can be estimated in second-order perturbation theory as
\begin{equation}
J = \frac{t_1^2}{U_0}\approx 0.5 \, \text{meV},
\end{equation}
where $t_1 = 0.022$ eV is the inter-trimer hopping energy, and $U_0 \approx 1$ eV is the estimated on-site Hubbard Coulomb repulsion energy for Frenkel excitons in Nb ions. 

Since a spin-1 particle on a triangular lattice is expected to behave more classically than a spin-1/2 particle, we propose to use the Weiss mean-field approximation for the spin-1 Frenkel excitons, in which the exchange interaction between neighboring spins can be represented as an effective exchange field acting on each spin. Specifically, each spin experiences a mean field due to its six neighboring spins, which effectively adds a Zeeman-like term to the Hamiltonian. We obtain a $120^{\circ}$ spin configuration for the spin-1 Frenkel excitons as the ground state, shown in Fig.~\ref{fig:excitonic_hubbard_model}. Note that the unit cell of this phase contains three sites.

Considering the exchange interaction \( J \) and applying the Weiss mean-field theory, the exchange term can be approximated as
\begin{equation}
J \sum_{\langle i,j \rangle} \mathbf{S}_i \cdot \mathbf{S}_j \approx M_{ex} \sum_i S_{i},
\end{equation}
where $M_{ex}=J\sum_{j=1}^6 \left<S_{j}\right>=\frac{J z S}{2}=3J$ is the mean field experienced by the spin-1 particle on each site, \( z=6 \) is the coordination number (number of nearest neighbors), and $S=1$ is the spin of the Frenkel exciton. Compared with the spin-1/2 system on a triangular lattice, the spin-1 system is more stable because the mean exchange field is twice as large.
Thus, in the mean-field approximation, the exchange interaction reduces to an effective Zeeman term, which is consistent with the spin splitting between the CB and VB obtained in the spin-DFT and noncollinear DFT calculations.

Our result is in agreement with recent experimental data suggesting antiferromagnetic coupling with a Weiss temperature of $T_W=-18.9$ K or $T_W=-13.1$ K in Nb$_3$Cl$_8$.\cite{Sun2022,Haraguchi2024} Its magnitude corresponds to an energy of about $|k_BT_W|=1$ meV, which is approximately of the order of magnitudes of the estimated values of $J$ and $D$. The Weiss temperature can be approximated by the formula $k_B T_W=-zJS(S+1)/3$ for a spin-$S$ system.\cite{Mugiraneza2022} Thus, we obtain $k_B T_W=-2J=-1$ meV, in excellent agreement with experimental data.
\subsection{Energy Hierarchy}
\textcolor{black}{Table~\ref{tab:energy_scales} summarizes the relevant energy scales, including the quasiparticle band gaps obtained from PBE and GW 
($E_{g}^{\mathrm{PBE}}$, $E_{g}^{\mathrm{GW}}$), the anisotropy energy of the dipole moment ($E_{D}^{\perp}-E_{D}^{\parallel}$), the dipole--dipole interaction energy (in-plane $E^{\parallel}_{DD}$ and out-of-plane $E^{\perp}_{DD}$), intra-trimer hopping $t_0$, inter-trimer hopping $t_1$ the on-site Hubbard repulsion ($U$), and the exchange interaction energy ($E_{\mathrm{exc}}$). From Table~\ref{tab:energy_scales}, it is evident that the out-of-plane  anisotropy energy of the electric dipoles exceeds the electric dipole--dipole interaction energy by nearly two orders of magnitude. This implies that all electric dipoles preferentially align perpendicular to the Nb$_3$Cl$_8$ plane in a single direction due to the mirror symmetry breaking of the Nb$_3$Cl$_8$ lattice in $z$ direction. Therefore, our single-cell Nb$_3$Cl$_8$ numerical calculations capture fully the ground state properties of the coupled electric dipoles.
Since the spin exchange interaction energy is much smaller than the electric dipole anisotropy energy, we can treat it as a correction to the DFT-GW-BSE calculations by increasing the unit cell by a factor of three to allow for the various possible spin configurations on the BKL, as described in Sec.~\ref{sec:magnetic_ordering}.}

\begin{figure}[!t]\includegraphics[width=3.2in]{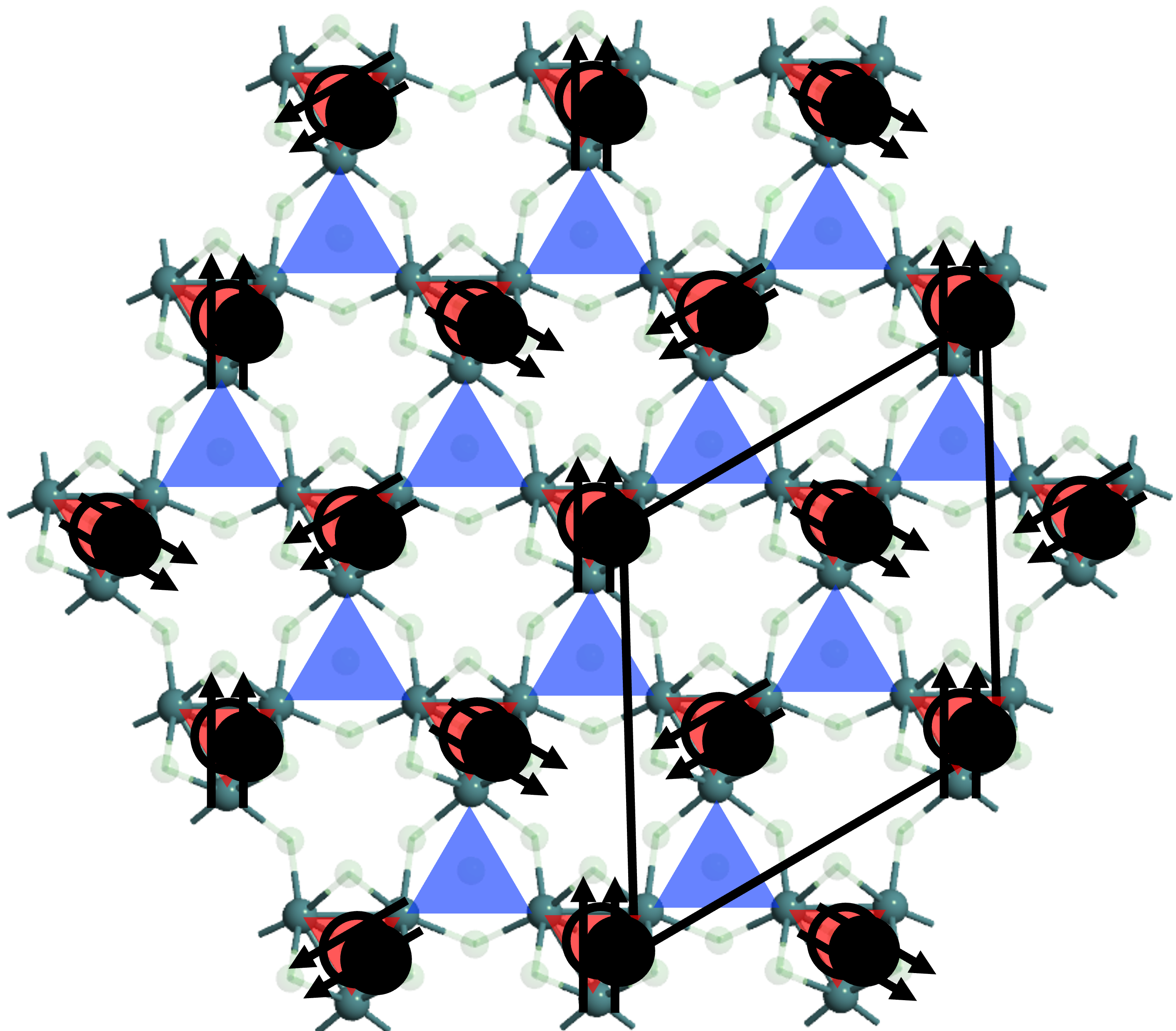}
\centering
\caption{ Each trimer of Nb atoms (dark green balls) has a spin triplet exciton of total spin 1. i.e. both electron (solid black circle) and hole (overlapping empty circle) has parallel in-plane intrinsic spins.}
\label{fig:excitonic_hubbard_model}
\end{figure}

\begin{table}[t]
\centering
\caption{Comparison of relevant energy scales (eV) in Nb$_3$Cl$_8$.}
\label{tab:energy_scales}
\begin{tabular}{|c| c| c| c| c| c| c| c| c|}
\hline
$E_{g}^{\mathrm{PBE}}$ & 
$E_{g}^{\mathrm{GW}}$ & 
$E_{D}^{\perp}-E_{D}^{\parallel}$ & 
$E_{DD}^{\parallel}$ & 
$E_{DD}^{\perp}$ &
$t_0$&
$t_1$&
$U$ & 
$E_{\mathrm{exc}}$ \\
\hline
0.27 & 2.27 & -7.08 & -0.0087 & 0.0258 & -0.325 & 0.022 & 1 & 0.1 \\
\hline
\end{tabular}
\end{table}

\section{Discussion}
The ab-initio GW-BSE calculation for SL Nb$_3$Cl$_8$ reveals two important features: 
(i) The lowest-energy exciton is dark due to the spin-triplet configuration of the electron-hole pair. This exciton lies at E$_{\rm dark}=-$0.14 eV and has a binding energy of 2.64 eV, indicating an exciton insulator state.\cite{Gao2023,Haraguchi2024} 
(ii) The bright excitons lie at energies 0.93~eV and 1.2~eV, with large binding energies of 2.05~eV and 1.77~eV, respectively. This substantial binding is attributed to the reduced dimensionality and the presence of flat bands, which enhance Coulomb interactions and suppress screening. These characteristics make SL Nb$_3$Cl$_8$ a promising platform for realizing strongly bound excitonic states and exploring correlated exciton physics.

Our \textit{ab initio} GW--BSE results refine the understanding of the ground-state quantum phase of Nb$_3$Cl$_8$. While previous theoretical studies successfully captured many features of the bulk material using an electronic spin-1/2 Hubbard-model framework, where moderate electronic correlations open a Mott gap near 1~eV and reproduce the observed optical transition, our findings indicate that this picture becomes less adequate in the strictly two-dimensional limit. 
In the single-layer regime, the substantial reduction in dielectric screening strongly enhances Coulomb interactions, pushing the system beyond the conventional electronic Mott regime. 
Consequently, the quasiparticle energies are renormalized below the GW band gap, leading to the emergence of a dark excitonic state at $-0.14$~eV with an exceptionally large binding energy of 2.64~eV. This state corresponds to a tightly bound spin-triplet Frenkel exciton localized on each Nb trimer, establishing the single-layer phase as an excitonic Mott insulator. 
In this regime, the low-energy physics is governed not by electron hopping between Hubbard bands but by the collective dynamics of interacting excitons, marking a crossover from a bulk-like electronic Mott state to a two-dimensional dark excitonic phase.

Using the classical electric dipole-dipole interaction, we estimate that the electric dipoles of the Frenkel excitons to orient in ferroelectric out-of-plane configuration.
In contrast to conventional magnetoelectric multiferroic materials, where for example charge transfer emerges when electron redistribution breaks spatial inversion (leading to polarization) while simultaneously mediating spin interactions (leading to magnetism),\cite{Spaldin2005,Ramesh2007,Cheong2007} multiferroicity in Nb$_3$Cl$_8$ originates from spin-1 excitons. This excitonic origin points to a fundamentally different route to multiferroicity, driven by collective quasiparticle interactions rather than individual electronic charge transfers, offering new possibilities for tunable quantum phases in low-dimensional systems. Possible technological applications include transducers converting between magnetic and electric fields, attenuators, filters, field probes, and data recording devices based on electric control of magnetization and vice versa.\cite{Spaldin2005}

We predict that the combination of flat bands and electric dipole ordering should give rise to enhancement of second-harmonic generation (SHG), third-harmonic generation (THG), and high-harmonic generation (HHG) in Nb$_3$Cl$_8$, in particular because the antiferroelectric dipole configuration on a triangular lattice breaks the inversion symmetry. While SL Nb$_3$Cl$_8$ should exhibit nonlinear response for all orders $n$, i.e. nonzero $\chi^{(n)}$ for all integers $n>2$ due to inversion symmetry breaking, we anticipate that for two and in general an even number of AB-stacked layers of Nb$_3$Cl$_8$ the even orders of nonlinear response should vanish, because the inversion symmetry is restored. The reason is that AB stacked (or bulk) Nb$_3$Cl$_8$ exhibits D$_{3d}$ point group symmetry,\cite{jeff2023raman} which contains the inversion symmetry. Note that bulk Nb$_3$Cl$_8$ is typically AB-stacked at room temperature.\cite{Sheckelton2017}

\section{Acknowledgments}
M. N. L. acknowledges support by the Air Force Office of Scientific Research (AFOSR) under award no. FA9550-23-1-0455
M. N. L. and D. R. E. acknowledge support by the AFOSR under award no. FA9550-23-1-0472.
Calculations were performed at the Stokes high performance computer cluster of the University of Central Florida.
Some calculations were performed on the Darwin high performance computer cluster provided by the ACCESS program of the National Science Foundation (NSF).
M. N. L., D.S., and M. A. K. acknowledge support by the NSF ACCESS program under allocation no. PHY230182 and no. PHY240242.

\section{Methods}
 We consider a unit cell of SL Nb$_3$Cl$_8$, consisting of 11 atoms in a hexagonal structure, with edge lengths $a=b=6.81$ \AA. A vacuum layer of $20$~\AA~is introduced in the $z$-direction to minimize interactions between the periodic images of Nb$_3$Cl$_8$.
 
We first perform mean field DFT in generalized gradient approximation (GGA) as implemented in VASP,\cite{vasp} on a 9$\times$9$\times$1 k-grid with a cut-off energy of 400 eV. For PBE Calculations a band gap of 0.27 eV is obtained Fig.~\ref{fig:Nb3Cl8_Structure_Bs} (b) (dashed line). The obtained results are consistent with previous DFT calculations.\cite{Sun2022} We use the PBE non-collinear SOC results as the starting point for our GW calculations.

The GW calculations were initiated using DFT-derived wavefunctions. Self-consistency was achieved by iteratively updating the GW eigen-energies, thereby removing dependence on the single-particle energies obtained from the initial DFT computations.
For GW calculations we consider a 9$\times$9$\times$1 grid with a cut-off energy (ENCUT) of 400 eV. For response function the we set ENCUTGW$=$200 eV. The parameter ENCUTGW controls the basis set for the response functions in exactly the same manner as ENCUT for the wave functions. We use 324 empty bands,

Details of the convergence of conduction band, valence band and band gap with respect to the number of bands $N_b$ and number of iterations $N_I$ are shown in the Supplementary Information.
For BSE calculations, we consider 20 bands above and below the Fermi level, it is important to note that results seem to converge at 12 bands above and below the Fermi level.

Note that we use the results from the DFT-GW-BSE calculations as the starting point for Bose-Hubbard mean field calculations.

\section{Author Contributions}
M. A. K. provided theoretical insights, performed analytical calculations, performed and analyzed numerical calculations, and co-wrote the manuscript.  N.U.D. and D.S. performed and analyzed numerical calculations. D.R.E. provided experimental insights and assisted with manuscript revisions. M.N.L. supervised the overall research direction, provided theoretical insights, developed the theoretical framework, performed analytical calculations, and co-wrote the manuscript.  All authors discussed the results and reviewed the manuscript.

\section{Data availability}
The authors declare that the main data supporting the findings of this study are available within the paper and its Supplementary files. Part of the source data is provided in this paper. The data that support the findings of this study are available from the corresponding author upon reasonable request.

\section{Competing interests}
The authors declare no competing interests.
\bibliography{bib}
\end{document}